\documentclass{article}
\usepackage{authblk}
\usepackage{natbib}
\usepackage{color}
\usepackage{booktabs}
\usepackage{multirow}
\usepackage{amsmath}
\usepackage{amssymb}
\usepackage{rotating}
\usepackage{hyperref}
\usepackage[a4paper]{geometry}
\usepackage{graphicx}
\usepackage{siunitx}
\usepackage{etoolbox}           

\hypersetup{
    colorlinks=true,
    linkcolor=blue,
    filecolor=magenta,      
    urlcolor=cyan,
    citecolor=cyan,
  }

\renewcommand{\bfseries}{\fontseries{b}\selectfont} 
\robustify\bfseries             
\newrobustcmd{\B}{\bfseries}    

\newcommand{\justkd}{\ensuremath{\textrm{MMD}}}
\newcommand{\justfad}{\ensuremath{\textrm{FAD}}}

\newcommand{\mcs}[1]{\ensuremath{\textit{#1}}}

\newcommand{\aps}[3]{\ensuremath{\textrm{S}^{(#1)}_{#2}(#3)}}

\title{Measuring Audio Prompt Adherence with Distribution-based Embedding Distances\\ \vspace{1em}\large{Technical Report (March 2024)}}
\author{Maarten Grachten}
\affil{Sony Computer Science Laboratories, Paris, France}
\date{}

\begin{document}
\maketitle

\begin{abstract}
  An increasing number of generative music models can be conditioned on an audio prompt that serves as musical context for which the model is to create an accompaniment (often further specified using a text prompt).
  Evaluation of how well model outputs adhere to the audio prompt is often done in a model or problem specific manner, presumably because no generic evaluation method for audio prompt adherence has emerged.
  Such a method could be useful both in the development and training of new models, and to make performance comparable across models.
  In this paper we investigate whether commonly used distribution-based distances like Fréchet Audio Distance (FAD), can be used to measure audio prompt adherence.
  We propose a simple procedure based on a small number of constituents (an embedding model, a projection, an embedding distance, and a data fusion method), that we systematically assess using a baseline validation.
  In a follow-up experiment we test the sensitivity of the proposed audio adherence measure to pitch and time shift perturbations.
  The results show that the proposed measure is sensitive to such perturbations, even when the reference and candidate distributions are from different music collections.
  Although more experimentation is needed to answer unaddressed questions like the robustness of the measure to acoustic artifacts that do not affect the audio prompt adherence, the current results suggest that distribution-based embedding distances provide a viable way of measuring audio prompt adherence.
  An python/pytorch implementation of the proposed measure is publicly available as a github repository.
\end{abstract}

\section{Introduction}\label{sec:introduction}
With the success of generative approaches like generative adversarial networks (GANs) and diffusion models (DM), generative AI for music has rapidly become a research topic of major interest.
A variety of tools have been proposed to create musical audio, in the form of full mixes \citep{agostinelli2023musiclm,donahue2023singsong,Forsgren_Martiros_2022,schneider2023mousai},
loops \citep{roberts2019hierarchical},
individual instrument parts \citep{DBLP:conf/waspaa/Lattner19,grachten20:_bassn,wu2022jukedrummer,parker2024stemgen,pasini2024bass},
or individual instrument sounds (one-shots) \citep{engel2017neural,nistal2020drumgan}.

The most common form of user control over the generated audio is by text prompts \citep{agostinelli2023musiclm,schneider2023mousai,huang2023make,huang2023noise2music,Forsgren_Martiros_2022,liu2023audioldm}.
\cite{wu2023music} provide more precise control by time-varying controls like rhythmic and dynamic envelopes, and melodic lines. 
Another form of user control is by conditioning on audio signals.
The conditioning audio is typically used either in a \emph{style-transfer} task, or an \emph{accompaniment} task.
In the former, the objective to generate audio that reproduces specific aspects of the conditioning audio (like the melody, the timbre, or the rhythm).
In the latter, the objective is to generate audio that goes well with, or complements, the conditioning audio
 \citep{grachten20:_bassn,parker2024stemgen,pasini2024bass,donahue2023singsong}.

Quantitative evaluation of generative models for musical audio is often done using multiple evaluation criteria, depending on the task.
The most commonly used evaluation criterion is audio quality, which is typically measured by the \emph{Fréchet Audio Distance} (FAD) \citep{kilgour2019frechet}, or similar distances, like the \emph{Maximum Mean Discrepancy} (MMD) \citep{gretton2012kernel}. 
Both metrics were first used in the image domain, and measure quality by computing the distance between real and generated data distributions in an embedding-space.

Secondly, generated outputs are evaluated in relation to the inputs through the judgment of human subjects, often expressed as the \emph{Mean Opinion Score} (MOS) \citep{donahue2023singsong,parker2024stemgen}.
Depending on the questions to the subjects, this form of evaluation can measure a variety of aspects, including adherence of the outputs to any prompts, but due to the involvement of human subjects, is not as readily available as computational evaluation methods. 


Thirdly, for text-conditioned music generation, prompt alignment is commonly measured by computing similarity between embedding vectors of the prompt and generated output in a joint embedding space, such as the \emph{CLAP score} \citep{huang2023make}, or the \emph{MuLan Cycle Consistency} \citep{agostinelli2023musiclm}.

Models conditioned on audio or symbolic representations of music, have been evaluated by measuring some form of reconstruction accuracy of desired characteristics, like melody, harmony, or rhythm \citep{wu2023music}, or a comparison of input and output audio in terms of musical descriptors \citep{parker2024stemgen}.
The former type of evaluation is most appropriate in style-transfer settings, whereas the latter is oriented towards musical accompaniment, under that assumption that stems belong to the same stem have similar distributions of musical descriptors.

To our knowledge, there are currently no generic quantitative measures of audio prompt adherence that are both instrument-agnostic, and do not make music-specific assumptions.
We address this by proposing a distribution-based method to measure for audio prompt adherence.
The method is non-specific to instrumentation in either the conditioning audio or the output audio, and is based on the commonly used FAD and MMD measures.

We validate the prompt adherence measure using a baseline evaluation in which we test the ability of the measure to reliably discriminate between matching and non-matching pairs of audio prompt and audio target.
In this setting, we systematically compare the constituents of the measure, including the underlying distance metric, the embedding model, the fusion method for prompt and target, and the projection of embeddings into a lower dimensional space.

We identify the most promising constituent combinations, and show that measuring audio prompt adherence by directly applying the distance metric to fused embeddings works when the reference and candidate data are from the same music collection, but not when reference and candidate data originate from different music collections.
Based on this result, we formulate an alternative measure that is much more sensitive to differences in prompt adherence, even when comparing across music collections.

Following the baseline evaluation, evaluate the sensitivity of the proposed measure to artificial perturbations of prompt adherence, applying time and pitch shift operations to the audio targets.
A python implementation of the audio prompt adherence measure is publicly available at \url{https://github.com/SonyCSLParis/audio-metrics}.

\section{Related Work}
\label{sec:relwork}

\subsection{Measures for audio quality}\label{sec:meas-audio-qual}

\cite{kilgour2019frechet} proposed the Fréchet Audio Distance as a measure of audio quality in the context of music enhancement algorithms.
Rather than evaluating individual audio samples, it evaluates a set of samples as a whole.
It does so by comparing the statistics of embedding representations of the set to be evaluated (the \emph{candidate} set) to those of another set of samples---the \emph{reference} set.
Specifically, the embeddings are assumed to follow a multi-variate normal distribution, and are represented by the mean and covariance of the distribution.
\cite{kilgour2019frechet} use the activations of the last feature layer of the VGGish model  \citep{hershey2017cnn} as embeddings.
Recent experiments by \cite{gui2023adapting} using FAD in the context of music generation reveal that music specific embedding models, especially the CLAP joint text/audio embedding model \citep{elizalde2023clap}, are more effective than VGGish.

An analogous metric is the \emph{Kernel Inception Distance} \citep{sutherland2018demystifying}, which uses the squared \emph{Maximum Mean Discrepancy} (\justkd{}) to measure distances between distributions of Inception embeddings. \justkd{} \citep{gretton2012kernel} measures the distance between two samples of embeddings by mapping them to a reproducing kernel Hilbert space (RKHS) and computing the distances between the means of the distributions in the RKHS. For suitable kernels like polynomial or Gaussian kernels, the distance can be computed analytically because of the kernel trick. 
\cite{nistal2020comparing} use the KID/\justkd{} metric in the context of audio generation to measure faithfulness of generated outputs in terms of pitch and instrument, by using two Inception models for these tasks, respectively.

\subsection{Measures for text prompt adherence}\label{sec:measures-text-prompt}
For generative models of musical audio that are conditioned on text prompts, the degree of adherence of the generated audio to the text prompt has been measured using the  \emph{CLAP score}. This measure is defined as the (averaged) cosine similarity between embeddings of the audio and the text in a joint text/audio embedding space defined by the CLAP model \citep{elizalde2023clap}. The \emph{MuLan Cycle Consistency} is defined similarly \citep{agostinelli2023musiclm}, using the embeddings of the \emph{MuLan} model \citep{huang2022mulan}.

\subsection{Other evaluation measures for conditional music generation}
\label{sec:other-qual-meas}
An analogous approach to measuring text prompt adherence could work to measure audio prompt adherence, but would require a joint audio prompt/target embedding model.
No generic models prompt/target models have been published as of yet, to our knowledge.
However \cite{pasini2024bass} trained a joint mix/bass embedding model in a contrastive manner to measure adherence of generated bass tracks to input mixes by computing dot products in the joint embedding space.

\cite{wu2023music} condition a generative model on symbolic melodies, and time-varying dynamics and rhythm curves.
They measure output adherence to these prompts by extracting melody, dynamics and rhythm descriptors from the output audio, and compute accuracy, correlation, and F1 score, respectively.

\cite{parker2024stemgen} propose MIRDD, a measure for audio prompt adherence that computes the Kullback-Leibler divergence between two distributions of pitch-, rhythm-, and structural-based audio descriptors, where the reference distribution is computed from mixes of conditioning audio and \emph{target} stems, and the candidate distribution is computed from mixes of conditioning audio and \emph{generated} stems.

\section{Method}\label{sec:method}
In this Section we describe our overall approach, including the evaluation methodology, data collections used, and the constituents of the computational pipeline.

\subsection{Baseline evaluation}\label{sec:baseline-eval}
We have no a-priori access to groundtruth data for audio prompt adherence to assess the quality of metrics, but we can define a baseline evaluation using multitrack audio data.
A multitrack audio collection consists of a number of \emph{projects} (typically songs), each of which contains one or more \emph{stems}.
The stems are the waveforms of the individual instruments involved in a project.

We make the assumption that when a subset of stems from a project is mixed into a single waveform---by samplewise summation---and used as an audio prompt, then all of the remaining stems in that project adhere to that prompt (if they were regarded as candidate stems to complement the prompt).
Conversely, we assume that stems from any other project will not adhere to that prompt.
Together, these assumptions suggest a straight-forward baseline evaluation of metrics for audio prompt adherence, according to which \emph{the best metric is the one that maximizes the expected difference between non-matching prompt/stem pairs and matching prompt/stem pairs} respectively. 
In Experiment 1 (Section~\ref{sec:experiment-1}) we evaluate the FAD and \justkd{} against this baseline.

For later reference, we introduce the following notation.
Let $X = \{(p_1, s_1), \cdots, (p_N, s_N)\}$ be a set of matching prompt/stem pairs of constant length.
Let $S= \{s_1, \cdots, s_N\}$ be the set of stems occurring in $X$.
Then define a set $X'$ as a copy of $X$ where elements $s$ have been permuted to pair items $p$ and $s$ at random (avoiding coincidental matches). Formally: 

\begin{equation}
  X'= \{(p_i, s_j)\ |\ 1 \leq i \leq N,\ s_j\in S, \textit{ and } j \neq i\}.
\end{equation}
We refer to $X$ as a \emph{matching set} and $X'$ as a \emph{non-matching} set.
Furthermore, for two sets $X$ and $Y$, we will use $\mathcal{M}_X(Y)$ to denote the distance of $Y$ to $X$, where $\mathcal{M}$ is either FAD or MMD.

\subsection{Data collections}\label{sec:datasets}
\begin{table}[t]
  \small
  \centering
  \begin{tabular}{lrrr}
    \toprule
    Collection & \# songs & Avg. stems / project & total duration (h) \\
    \midrule
    MUSDB18 \citep{MUSDB18HQ}& 150   &  4.0 &    9.8 \\
    A (in-house)       & 19998 & 11.7 & 1255.4 \\
    B (in-house)       & 583   & 10.3 &   28.1 \\
    C (in-house)       & 124   & 16.0 &    8.4 \\
    D (in-house)       & 573   & 12.8 &    2.5 \\
    \bottomrule
  \end{tabular}
  \caption{Data collections}
  \label{tab:collections}
\end{table}

The multitrack collections used in this study are listed in Table~\ref{tab:collections}.
The MUSDB18~\citep{MUSDB18HQ} collection is publicly available and consists of pop/rock songs.
The other collections are licensed for internal use only.
Of these, A is by far the largest with almost 20k songs in a variety of genres.
The remaining collections are smaller and consist of production music (B, C) and trap sample packs (D).
Collection D consists mainly of short segments (loops) rather than full songs.

In some cases the components of drums and percussion (like kick drum, snare drum, and hi-hat) are stored as separate stems, rather than in a single stem.
This accounts for the difference in steps per project between e.g. C, and MUSDB18.
Note that instruments/components typically not active throughout the entire song, so stems are often partly silent.

For the experiments (Sections~\ref{sec:experiment-1} to \ref{sec:exp3:pitchtime}) we split the projects of each collection into a \emph{reference} dataset and a \emph{candidate} dataset of equal size.
In the case of MUSDB18 we use the predefined training/test partition as the reference/candidate partition, which has a 2:1 size ratio.

\subsection{Distance metrics}\label{sec:dist-metrics}
To measure distances between data distributions in the embedding space, we use both FAD and \justkd{}, discussed in Section~\ref{sec:meas-audio-qual}.
In this work we use the \justkd{} with a polynomial kernel of degree 3, with $\gamma = 1/d$, where $d$ is the dimensionality of the embedding space, and $\mathit{coef}_0 = 1$, leaving experimentation with other kernels/parameters as future work.

\subsection{Embedding models}\label{sec:embedd-models}
The FAD and \justkd{} metrics are distribution-based metrics in the sense that they compare the audio material to be evaluated against a distribution computed over a reference data set.
The distribution is not computed directly from the raw data representation (the audio waveform), but from an embedding space.
When FAD values are reported in the literature, they are typically computed on embeddings obtained from \emph{VGGish}---a convolutional neural network that was pretrained on an audio classification task \citep{hershey2017cnn}.
The VGGish model was not specifically trained to be sensitive to musical characteristics of the sound, but more generally to characteristics relevant to auditory scene classification/event detection.
That it has nevertheless been used extensively for evaluation in music generation tasks \citep{nistal2020drumgan,pasini2022musika,caspe2022ddx7}, is testimony to the flexibility and robustness of the FAD as a metric.
Nevertheless, as shown by \cite{gui2023adapting}, more reliable FAD scores for music generation can be achieved by using music specific embedding models, specifically CLAP \citep{wu23:_laionclap_large_scale_contr_languag_audio}.

In this study we extract audio embeddings using three different pre-trained models: VGGish \citep{hershey2017cnn}, OpenL3 \citep{cramer2019look}, and CLAP \citep{wu23:_laionclap_large_scale_contr_languag_audio}.
From VGGish we use the last feature layer (the layer before the classification layer).
OpenL3 outputs the embeddings themselves.
CLAP consists of both an audio and text encoder.
We only use the audio encoder for the present work.
Specifically, we use the output layer and the two preceding audio projection layers from the audio encoder as separate embeddings.
Table~\ref{tab:embedding-layers} summarizes the embeddings and their respective dimensionalities.
We will use the labels listed in the table to refer to the respective embeddings, and use $\theta(x): \mathbb{R}^T \rightarrow \mathbb{R}^D$ to denote a generic embedding function that maps an digital mono audio waveform of $T$ samples to an embedding vector of dimensionality $D$.

For VGGish we use the original checkpoint trained on AudioSet.\footnote{\url{https://github.com/harritaylor/torchvggish}}
For OpenL3 we use the ``mel256'' model trained on music.\footnote{\url{https://github.com/marl/openl3}}
For CLAP we use the pretrained model trained on music, provided by LAION as the checkpoint labeled \verb+music_audioset_epoch_15_esc_90.14.pt+.\footnote{\url{https://github.com/LAION-AI/CLAP}}. 

\begin{table}[t]
  \small
  \centering
  \begin{tabular}{llrc}
    \toprule
    Model & Layer ($\theta$) & \# dim & Label\\
    \midrule
    VGGish & Last feature layer &  128  & VGG \\
    OpenL3 & Output layer &  6144       & OL3 \\
    Laion CLAP & 1st audio projection &  512 & CLAP0\\
          & 2nd audio projection &  512           & CLAP1\\
          & Output layer &  128              & CLAP2 \\
    \bottomrule
  \end{tabular}
  \caption{Embedding models and layers used in this study}
  \label{tab:embedding-layers}
\end{table}

\subsection{PCA projection}\label{sec:pca-proj}
Table~\ref{tab:embedding-layers} shows that the dimensionality varies considerably across the embeddings.
In particular the dimensionality of OL3 is substantially larger than that of the others.
Since distribution-based metrics like FAD and \justkd{} essentially measure the overlap between distributions, high dimensional spaces---which are more likely to be sparse---may decrease the effectiveness of the metrics.
To test the benefit of lower dimensional embeddings, if any, we consider two whitening PCA projections of the embeddings, in addition to using the original embeddings, as listed in Table~\ref{tab:pca-options}.
We will refer to the projection operation (which may be the identity function) as $\gamma$.

\begin{table}[t]
  \small
  \centering
  \begin{tabular}{lc}
    \toprule
    Projection ($\gamma$) & Label \\
    \midrule
    No projection (Identity)     & NP \\
    PCA 10 components  & PCA10 \\
    PCA 100 components & PCA100 \\
    \bottomrule
  \end{tabular}
  \caption{Projection options considered in this study}
  \label{tab:pca-options}
\end{table}

\subsection{Prompt/stem fusion method}\label{sec:fusion-methods}
In order to use the FAD and \justkd{} metrics to quantify the relationship between a stem and a prompt, the stem and prompt must be jointly represented as a single vector.
The most straight-forward method of performing this fusion operation is to mix the prompt and stem waveforms into a single waveform and compute the embeddings for the mix.
Alternatively, the embeddings can be computed for prompt and stem individually, and them combined, for example by summing, or by concatenation.
We evaluate these three fusion methods, listed in Table~\ref{tab:fusion-options}, in the experiments.
The last column specifies for each fusion method how the embedding $\theta$, and projection $\gamma$ are used to define the function $\phi$ that maps audio pairs $(p, s)$ to the space on which the metrics are computed.
In the remainder of this work we use $\phi_X = \{ \phi(p, s)\ |\ (p, s) \in X \}$ to denote the set of embeddings computed from the pairs $(p, s)$ in $X$.

\begin{table}[t]
  \small
  \centering
  \begin{tabular}{lccl}
    \toprule
    Prompt/Stem fusion method & Fusion stage & Label & Pipeline $\phi(p, s)\!:\ \mathbb{R}^T\! \times \mathbb{R}^T \rightarrow \mathbb{R}^D$\\
    \midrule
    Mix audio waveforms     & Early& MIX & $ \phi(p, s) =  \gamma (\theta (p + s) ) $ \\
    Sum embeddings          & Late & SUM & $ \phi(p, s) =  \gamma (\theta (p) + \theta(s) ) $ \\
    Concatenate embeddings  & Late & CONC& $ \phi(p, s) =  \gamma ([\theta (p), \theta(s) ]) $ \\
    \bottomrule
  \end{tabular}
  \caption{Fusion methods for prompt and stem considered in this study. In the last column $\theta$ and $\gamma$ represent the embedding model and projection respectively, $+$ represents elementwise summation, and $[\theta (p), \theta(s) ]$ represents the concatenation of embeddings $\theta (p)$ and $\theta(s)$ along the embedding dimension.}
  \label{tab:fusion-options}
\end{table}

\section{Experiment 1: Baseline exploration}\label{sec:experiment-1}
The purpose of this experiment is to test whether the FAD and KID metrics, using the data collections and the various configurations described in Section~\ref{sec:method}, are capable of discriminating non-matching from matching prompt/stem pairs.
Rather than comparing full length waveforms, we are interested in estimates of prompt adherence at---informally speaking---the shortest meaningful time scale.
Although certain acoustic characteristics possibly relevant to prompt adherence judgments (like genre-specific timbres) may be perceptible even in short timespans like 1s or less, most likely a slightly longer temporal context is necessary for a more comprehensive prompt adherence estimation, for example to take into account rhythmic characteristics.
Throughout the current work we use 5s windows as the basis for prompt adherence estimation.

\subsection{Procedure}
\label{sec:exp1-design}
From the reference set of each collection we sample 10000 time windows, and for each window we construct a prompt/stem pair (see Section~\ref{sec:exp1-data-processing}), yielding the \emph{matching reference set} ($A$).
We do the same for the candidate set of each collection, thus constructing the \emph{matching candidate set} ($B$).
From this set, we create the \emph{non-matching candidate set} ($B'$) as described in Section~\ref{sec:baseline-eval}.

To test the sensitivity of the FAD and \justkd{} metrics to prompt adherence, we use the matching reference set $A_i$ of collection $i$ to compute background distributions, and compare both the matching candidate set $B_j$ and the non-matching candidate set $B'_j$ of each collection $j$ to those background distributions.
We perform paired difference tests to test for a significant difference in the metric scores for $B_j$ and $B'_j$.
We use the non-parametric sign test to avoid normality and symmetry assumptions on the distribution of the differences.
The lack of statistical power of this test compared to tests like the paired t-test is not a problem, since we use the test as an objective criterion to establish a metric for prompt adherence, and we are not interested in very weak statistical effects.
For the same reason we deem the relatively small sample sizes of metric values ($n=5$ for within-collection comparisons, and $n=20$ for between-collection comparisons) unproblematic.

\subsection{Data processing}\label{sec:exp1-data-processing}
Prompt/stem pairs are obtained by sampling a 5s window from a project (with a hop size of 1s), designating one of the stems as the target stem, and creating a mix from a random subset of the remaining tracks to form the prompt.
Since large portions of stems can be silent, there is a considerable possibility that there is no signal in the 5s stem or prompt windows.
To avoid this we perform silence detection on the stems, and only select a window if there are at least two stems that are non-silent at the center of the window. 
This way we can ensure that both the prompt and stem are non-silent in all sampled windows.

\subsection{Results}
\label{sec:exp1-results}
The variance in the embeddings explained by the PCA10 and PCA100 subspaces is listed in Table~\ref{tab:explained-variance}.
In most cases PCA10 captures 60\% of the variance or more, whereas PCA100 typically captures 90\% or more.
As expected, the explained variance is partially related to the dimensionality of the embeddings: The explained variance is relatively low for the concatenated embeddings (CONC) and also for OL3.

\begin{table}[hbt!]
  \small
  \centering
  \begin{tabular}{llrr}
    \toprule
 &  & \multicolumn{2}{c}{Explained variance; mean (std)} \\
\cmidrule(rr){3-4}
Embedding  & Fusion method &   PCA10 & PCA100 \\
\midrule
    VGG   &  MIX & 0.75 (0.02)   & 0.99 (0.00) \\
    OL3   &  MIX & 0.62 (0.02)   & 0.86 (0.01) \\
    CLAP0 &  MIX & 0.58 (0.02) & 0.95 (0.00) \\
    CLAP1 &  MIX & 0.61 (0.02) & 0.97 (0.00) \\
    CLAP2 &  MIX & 0.61 (0.02) & 0.97 (0.00) \\
    \midrule
VGG   & SUM & 0.73 (0.04) & 0.99 (0.00) \\
OL3   & SUM & 0.66 (0.02) & 0.89 (0.01) \\
CLAP0 & SUM & 0.57 (0.04) & 0.95 (0.01) \\
CLAP1 & SUM & 0.58 (0.03) & 0.97 (0.00) \\
CLAP2 & SUM & 0.58 (0.03) & 0.97 (0.00) \\
    \midrule
VGG   & CONC & 0.57 (0.04) & 0.95 (0.01) \\
OL3   & CONC & 0.54 (0.03) & 0.85 (0.01) \\
CLAP0 & CONC & 0.44 (0.04) & 0.88 (0.01) \\
CLAP1 & CONC & 0.44 (0.04) & 0.91 (0.00) \\
CLAP2 & CONC & 0.44 (0.04) & 0.91 (0.01) \\
    \bottomrule
  \end{tabular}
  \caption{Explained variance of PCA projections for each embedding, computed from and averaged
    over reference sets of the music collections}
  \label{tab:explained-variance}
\end{table}

\begin{figure}[ht!]
  \centering
  \includegraphics[width=\textwidth]{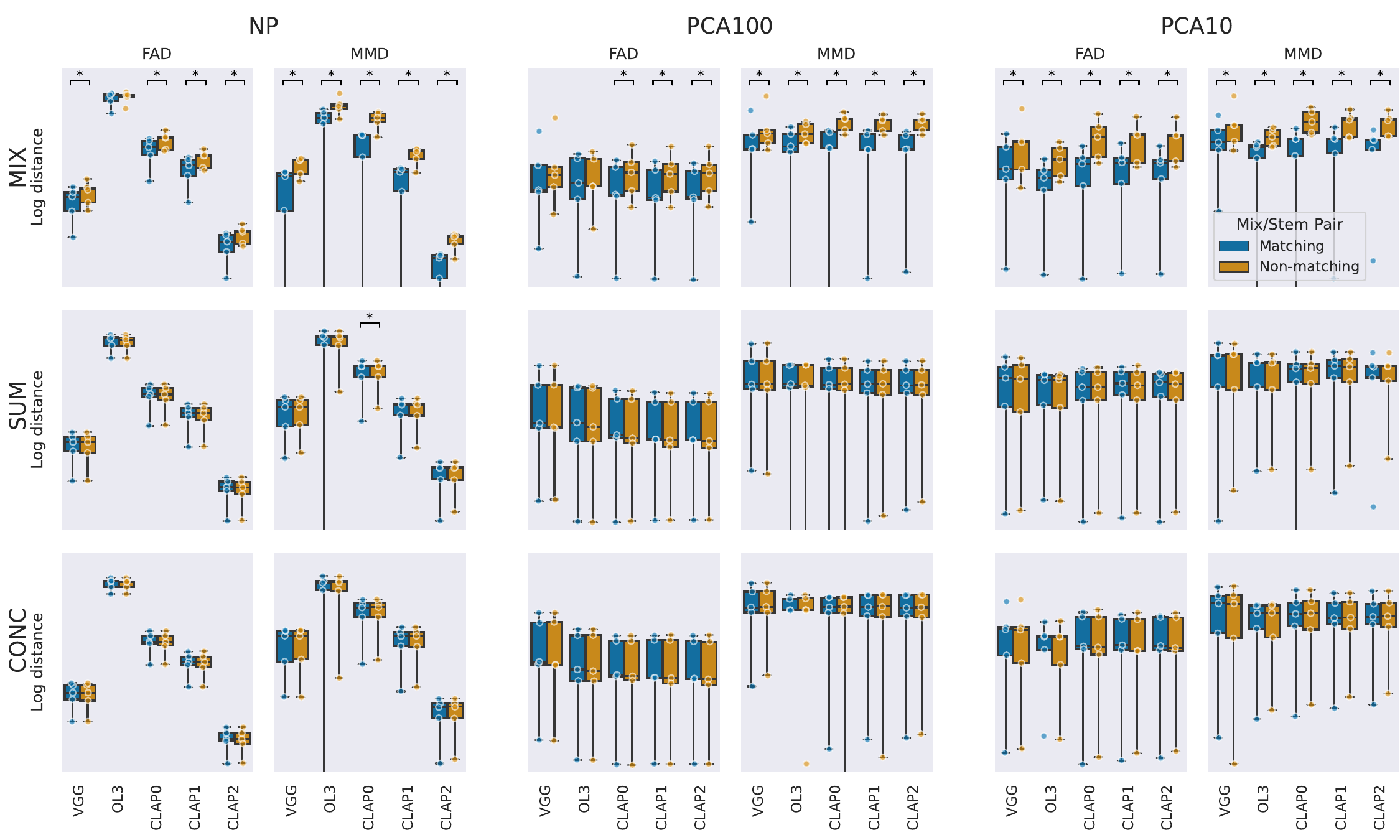}
  \includegraphics[width=\textwidth]{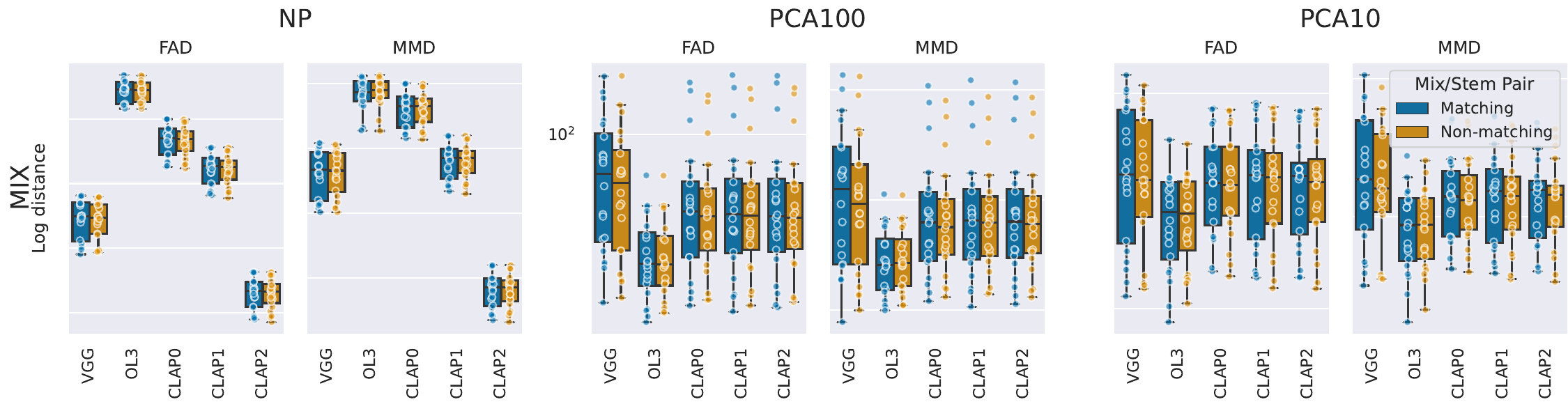}
  \caption{Within-collection (upper three rows) and between-collection (bottom row) FAD/\justkd{} distances of candidate sets to reference sets using different fusion methods, embedders, and projections. For between-collection distances, only the MIX fusion method is shown. The mix/stem pairs of the candidate sets are either \emph{matching} (blue) or \emph{non-matching} (orange).  Asterisks denote the statistical significance of differences.
  }
  \label{fig:boxplots-exp1}
\end{figure}

Figure \ref{fig:boxplots-exp1} shows the FAD/\justkd{} scores of matching and non-matching candidate sets ($B$ and $B'$ respectively), given the background distribution computed from the matching reference set ($A$), for all combinations of embedding model, projection, and fusion mode.
The upper three rows show results for \emph{within}-collection comparisons---where $B$ and $B'$ are from the same collection as $A$---using different fusion methods.
Asterisks denote statistical significance for rejecting the null hypothesis in favor of $\mathcal{M}_{A}(B') > \mathcal{M}_{A}(B)$ in the sign test.
Note that the significance of the sign test is limited by the small sample size for within-collection comparisons.
Specifically, the $p$-value has a lower bound of $\frac{1}{2^5} = 0.03125$, which means that results can be significant at most at the $\alpha=0.05$ level (single asterisk).

The within-collection comparisons in Figure~\ref{fig:boxplots-exp1} show that only the MIX fusion method leads to significantly different distances of $B$ and $B'$ to $A$, implying that the SUM and CONC methods are not effective fusion methods to measure prompt adherence.
Thus, for brevity, we only include between-collection distances for MIX in Figure~\ref{fig:boxplots-exp1} (bottom row).

\subsection{Discussion}
\label{sec:exp1-discussion}
The results show that the choice of the fusion method is crucial form measuring differences between matching and non-matching prompt/stem pairs using the tested embeddings.
The MIX fusion method computes a single embedding on the mixed prompt/stem waveform (early fusion), rather than summing or concatenating seperately computed prompt/stem embeddings (late fusion).
This suggests that the discriminative properties we are looking for arises from the embedding models ``being aware of'', and actively encoding the discrepancy between prompt and stem, and not purely by a divergence in the embedding space. 

Despite the MIX fusion method showing promise for within-collection comparisons, between-collection comparisons do not show lower distances for matching than for non-matching sets. 
To the contrary, the non-matching candidate set is sometimes even slightly closer to the (matching) reference set than the matching candidate set.
Even if this may appear counter-intuitive and discouraging at first, a simple geometrical interpretation suggests that the FAD and MMD may still serve as the basis for a audio prompt-adherence measure across collections.

\begin{figure}[t]
  \centering
  \includegraphics[width=.6\textwidth]{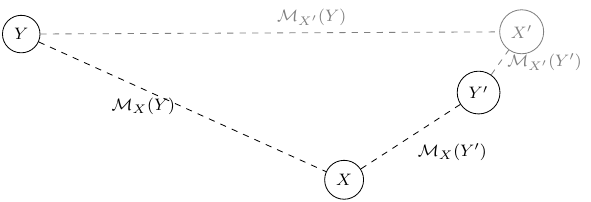}
  \caption{Hypothetical constellation of matching and non-matching reference ($X$, $X'$) and candidate sets ($Y$, $Y'$) for between-collection comparisons}
  \label{fig:constellation}
\end{figure}

To illustrate this, Figure~\ref{fig:constellation} shows a hypothetical constellation of matching and non-matching reference ($X$, $X'$) and candidate sets ($Y$, $Y'$) for between-collection comparisons.
In this setting we have that $\mathcal{M}_{\mcs{X}}(\mcs{Y}\, ') < \mathcal{M}_{\mcs{X}}(\mcs{Y})$.
However, $Y'$ here is not only closer to $X$ than $Y$, but also to the non-matching reference set $X'$, and is in fact closer to $X'$ than to $X$, whereas $Y$ is closer to $X$ than to $X'$.
This hypothetical constellation of datasets suggests that, rather than considering only the \emph{absolute proximity} of the candidate set and matching reference set $X$ (as in the current experiment), the \emph{relative proximity} of a candidate set to $X$ and $X'$ may be more informative for prompt-adherence.
In Experiment 2 we propose and evaluate a measure of prompt-adherence that incorporates this idea.
From now on we will focus only on the MIX fusion method, omitting CONC and SUM.

\section{Experiment 2: Baseline validation}\label{sec:experiment-2}
The results from Experiment 1 show that even if the metrics are sensitive enough to register distribution differences between matching and non-matching prompt/stem pairs, these differences are small, and only statistically significant if the there are no confounding factors like the reference and candidate sets originating from different music collections.
This is undesirable, since it restricts our ability to measure audio prompt adherence to cases where we can assume the reference and candidate sets are sufficiently similar.

As hinted at in the previous section, we can define an alternative measure of prompt adherence by comparing a set $Y$ to both the original reference set $X$, and a non-matching version $X'$ derived from $X$. 
Specifically, when $\mathcal{M}$ is a metric ($\justfad{}$ or $\justkd$), $X$ is a reference set of matching prompt/stem pairs, and $Y$ is the candidate set to compare against $X$, we propose the following score to measure prompt adherence of $Y$ relative to $X$:

\begin{align}
  \aps{\mathcal{M}}{X}{Y} = \frac{\mathcal{M}_{\mcs{X}'}(\mcs{Y}) -\mathcal{M}_{\mcs{X}}(\mcs{Y})}{\mathcal{M}_{\mcs{X}'}(\mcs{Y}) + \mathcal{M}_{\mcs{X}}(\mcs{Y})},
  \label{eq:prompt-adherence-score}
\end{align}

\noindent where $X'$ is a non-matching set constructed from $X$ by randomly pairing stems and mixes (cf. Section~\ref{sec:baseline-eval}).
To see how this addresses the issue of between-collection comparisons, note that $S$ measures the difference between two distances, namely $Y$ to $X'$ and $Y$ to $X$, normalized by the sum of both distances.
This means that the absolute values of the two distances are no longer relevant, only their proportion.
Furthermore, since $\mathcal{M}$ is a metric, and thus non-negative, $\aps{\mathcal{M}}{X}{Y}$ ranges between -1 and 1 for any sets $X$, $Y$ of prompt/stem pairs, provided $\mathcal{M}_{\mcs{X}'}(\mcs{Y})$ and $\mathcal{M}_{\mcs{X}}(\mcs{Y})$ are not both zero, in which case $\aps{\mathcal{M}}{X}{Y}$ is undefined.

Analog to the previous experiment, we now test the ability of $S$ to discriminate matching candidate sets $B$ from the non-matching candidate sets $B'$, especially for between-collection comparisons.
As in Experiment 1, sign tests are used to identify which configurations yield significant paired-sample differences between the matching and non-matching conditions.

\begin{figure}[t]
  \centering
  \includegraphics[width=\textwidth]{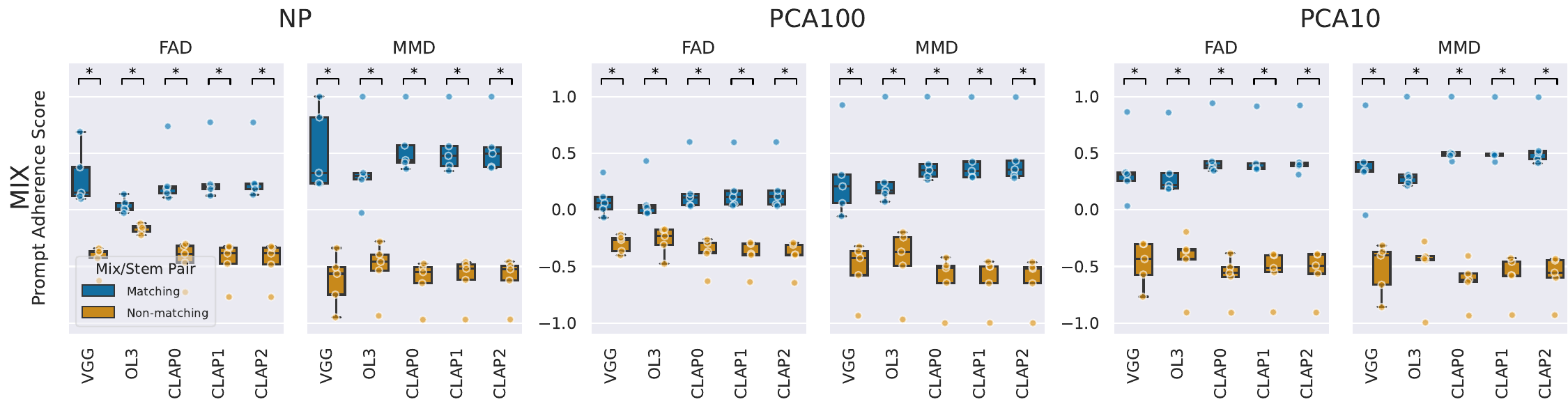}
  \includegraphics[width=\textwidth]{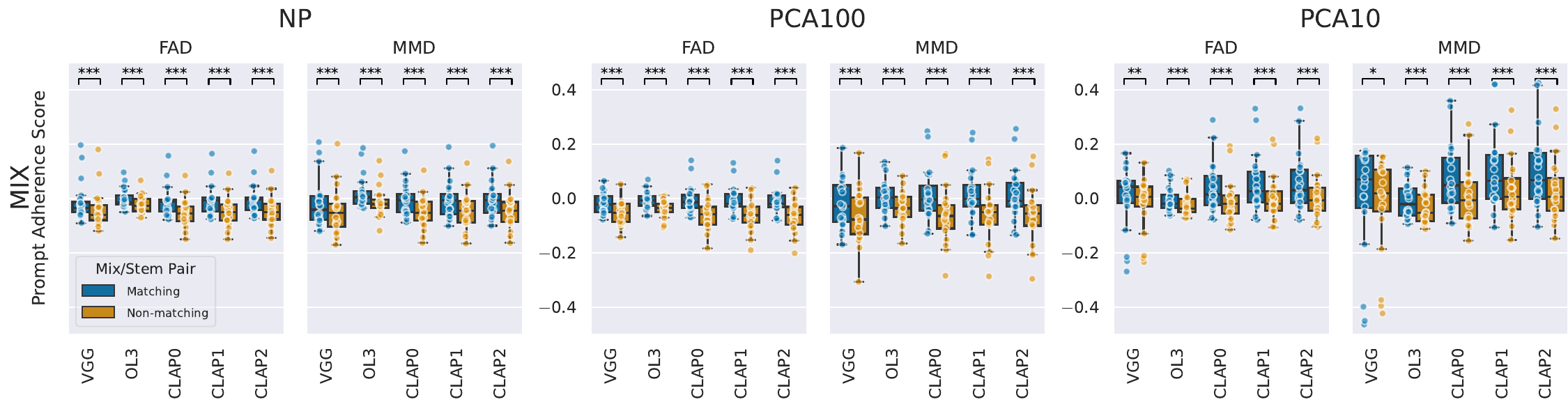}
  \caption{[Experiment 2] Within-collection (top) and between-collection (bottom) FAD/\justkd{}-based prompt adherence scores $S^{(\mathcal{M})}$
    of matching vs non-matching candidate sets}
  \label{fig:boxplots-exp2}
\end{figure}

Figure~\ref{fig:boxplots-exp2} shows the results for within- and between-collection comparisons using $S^{(\mathcal{M})}$, with asterisks indicating the significance of paired-sample differences, if any.
Even if the difference in scores is still much larger for within-collection comparisons than for between-collection comparisons, the sign-tests reveal that even for between-collection comparisons, the paired-sample differences are highly significant, regardless of the underlying metric, the projection method, and---surprisingly---even the embedding model to a large extent.

\section{Experiment 3: Pitch/time-shift sensitivity}\label{sec:exp3:pitchtime}
So far we have focused on a baseline evaluation method in which we test the ability of audio prompt adherence measure to distinguish between matching mix/stem pairs and randomly paired mixes and stems.
This has allowed us to rule out SUM and CONC as viable fusion methods, and has shown that direct use of the FAD and MMD distances to measure prompt adherence is not robust to differences between music from different collections.

To be of practical use however, the measure for audio prompt adherence should not just be able to distinguish between the extremes of the scale, but also be sensitive to intermediate forms of prompt adherence.
To test this we run the experiment with different non-matching conditions, in which we apply random pitch-shifting, and time-shifting to the stems of the mix/stem pairs.
We expect an intuitive measure audio prompt adherence to have the following properties:
\begin{description}
\item[h1] Scores are substantially lower for pairs where the target stems have been pitch or time shifted than for matching pairs
\item[h2] Scores are lower when both pitch and time shift operations have been applied than when only a single operation has been applied
\item[h3] The random mix/stem association condition produces lowest scores, with similar or slightly higher scores for the pitch + time shift condition
\end{description}
Audio prompt adherence scores should be significantly lower for pairs where the target stems have been pitch and/or time shifted that for matching pairs, but we expect the scores to 
Furthermore, we expect a useful measure to assign lower scores when both pitch- and time-shifts have been applied than when a single operation has been applied.


To create the non-matching mix/stem pairs, we apply random pitch-shifting of the stem by plus/minus 1 to 7 semitones (inclusive), and random time-shifting by plus or minus 0.2 to 2.5s.
We run the experimental setup of Experiment 2 with three different non-matching conditions: pitch-shift only, time-shift only, and simultaneous pitch- and time-shift.

The sign tests show highly significant paired-sample differences between matching and non-matching for all conditions, with slightly weaker results for VGG/OL3, and generally for PCA10.
This means that for a given reference set $A$ and candidate set $B$ (regardless whether $A$ and $B$ are from the same collection or not), when we derive a candidate set $B'$ from $B$ by pitch- and/or time-shifting the stems in $B$, then the audio prompt adherence score $S_A(B')$ will be significantly lower than $S_A(B')$.

\begin{figure}[t]
  \centering
  \includegraphics[width=\textwidth]{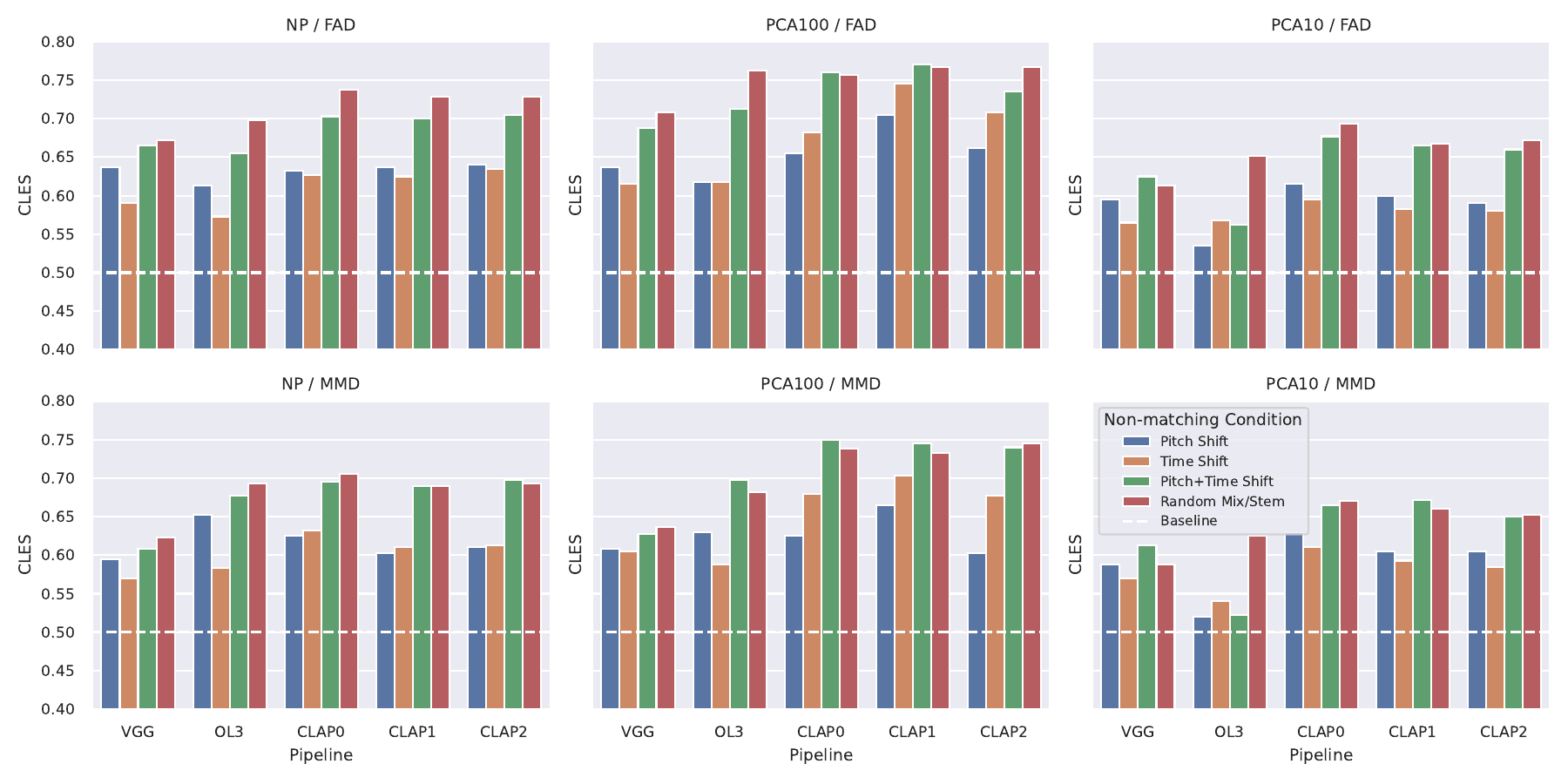}
  \caption{[Experiment 3] Common language effect size (CLES) of different non-matching conditions on audio prompt adherence score.
    }
  \label{fig:boxplots-exp3-between}
\end{figure}

Figure~\ref{fig:boxplots-exp3-between} plots the \emph{common language effect size} (CLES) \citep{mcgraw1992common} of the various non-matching conditions on unpaired samples. The (unpaired-sample) CLES value of a condition expresses effect size as the probability that given a reference set $A$, a randomly selected (matching) candidate set $B'$ to which the non-matching operation was applied, has a lower score than a second randomly selected candidate set ($C$) with matching mix/stem pairs:

\begin{equation}
\mathrm{CLES} = P(S_A^{(\mathcal{M})}(B') < S_A^{(\mathcal{M})}(C)) 
\label{eq:cles}  
\end{equation}

The null-hypothesis/baseline of the sign-test corresponds to a CLES value of 0.5, marked with a dashed white line in the plots.
Note that the \emph{higher} the CLES values, the larger the \emph{reduction} in audio prompt adherence score $S_A^{(\mathcal{M})}$ as a consequence of the non-matching condition. In other words, the CLES values reflect the sensitivity of the audio prompt adherence score to the non-matching conditions.

The plots show that generally the largest reductions in audio prompt adherence score are associated with the random mix/stem associations, and the pitch + time shifting operations, corroborating \textbf{h3}.
Furthermore, CLES values are generally higher for CLAP embeddings that for VGGish and OpenL3 embeddings.
Finally, the highest CLES values are obtained for PCA100 projections.
It is notable that PCA100 projections yield higher CLES values than the original embeddings.
This may be attributed to one or more properties of the whitening PCA projection.
For example, the ``compression'' of the data distributions into a lower-dimensional space may increase the overlap of reference and candidate distributions, leading to more meaningful FAD/\justkd{} distance values than when the distributions are mostly disjunct.
It is also possible that the whitening operation of the PCA enhances variance related to audio prompt-adherence, thus making the FAD/\justkd{} values more sensitive to these variations.
However, the lower CLES scores for the PCA10 projections show there is a limit to the beneficial effect of dimensionality reduction, suggesting that relevant details are lost in the highly reductive PCA10 projections.

Overall, the highest CLES values across the different non-matching conditions are obtained with FAD as the underlying distance, computed on PCA100 projections of the penultimate layer CLAP embeddings (CLAP1).

\section{Conclusion}
\label{sec:conc}
An increasing number of AI-based music generation systems allow for conditioning the generation process not only on a text prompt, but also on a musical audio context (or audio prompt).
As of yet there is no commonly accepted method to evaluate the quality of such systems in terms of how well generated musical outputs adhere to the audio prompt.
The aim of this report was to investigate the feasibility of---and lay the groundwork for---a measure for audio prompt adherence based on distribution based distances like FAD and \justkd{}.

We designed a pipeline with multiple constituents (distance, embedding model, embedding projection, and data fusion method) which we systematically evaluate in three consecutive experiments.
In Experiment 1 whether the distances (FAD and \justkd{}) can be used directly to measure audio prompt adherence. Although the results of the experiment show that this is not generally the case, interpretation of the results does lead to the formulation of an alternative measure that we refer to as the \emph{audio prompt adherence}  score (Equation~(\ref{eq:prompt-adherence-score}). In Experiment 2 we test this measure, and show that it is effective in distinguishing matching mix/stem pairs from randomly paired mixes and stems, even when the reference and candidate set are from different music collections.
In Experiment 3 we perform additional evaluation of the \emph{audio prompt adherence} measure, assessing its sensitivity to pitch and time shifting of the stems (while keeping mixes as is). Again we evaluate different combination of pipeline constituents.

Summarizing the findings across the experiments, we find that:
\begin{itemize}
\item The only fusion method that allows for measuring audio prompt adherence effectively is mixing the audio prompt with the output audio (early fusion), as opposed to computing embeddings separately and combining them (late fusion).
\item VGGish and OpenL3 embeddings generally are less effective than CLAP embeddings
\item FAD/\justkd{} distances---computed from embeddings of the mixed audio prompt and audio output---are not by themselves sensitive enough to reliably measure audio prompt adherence.
\item The proposed \emph{audio prompt adherence} score is sensitive to the tested non-matching conditions (random pairing, pitch shifting, and time shifting), both for comparisons within and across music collections. Absolute score values are somewhat collection specific, and are generally lower for across collection comparisons.
\item Overall, the highest sensitivity of the \emph{audio prompt adherence} score across the different non-matching conditions are obtained with FAD as the underlying distance, computed on PCA100 projections of the penultimate layer CLAP embeddings (CLAP1).
\end{itemize}

The preliminary experimentation reported here shows that the proposed measure meets some basic expectations of an audio prompt adherence score. Nevertheless, more work is needed to establish it as a reliable objective measure for real world application.
Next steps include demonstrating the insensitivity of the score to a number of audio quality degradations, which should not have a significant impact on audio prompt adherence.
In particular, since most state-of-the-art music generation approaches are a combination of an audio encoder/decoder with generation in the encoded/latent space, the measure needs to be tested for insensitivity to decoding artifacts---which degrade audio quality, but not audio prompt adherence.
Further work will also include a systematic assessment of the effect of relative loudness of audio prompt and output when mixing, as well as the effect of the sample size on the audio adherence score.

\bibliographystyle{apalike}
\bibliography{main-arxiv}
\end{document}